\documentstyle[12pt]{article}
\def\hybrid{\topmargin -20pt    \oddsidemargin 0pt
        \headheight 0pt \headsep 0pt
        \textwidth 6.25in       
        \textheight 9.5in       
        \marginparwidth .875in
        \parskip 5pt plus 1pt   \jot = 1.5ex}

\hybrid
\newskip\humongous \humongous=0pt plus 1000pt minus 1000pt
\def\caja{\mathsurround=0pt}
\def\eqalign#1{\,\vcenter{\openup1\jot \caja
        \ialign{\strut \hfil$\displaystyle{##}$&$
        \displaystyle{{}##}$\hfil\crcr#1\crcr}}\,}
\newif\ifdtup

\relax

\def\be{\begin{equation}}
\def\ee{\end{equation}}
\def\ba{\begin{eqnarray}}
\def\ea{\end{eqnarray}}
\def\d{\partial}
\def\db{\bar{\partial}}
\def\G{\Gamma}
\def\S{\Sigma}

\def\ta{\theta_1}
\def\tbb{\theta_2}

\def\s{\sigma}

\def\gt{{\tilde g}}

\begin{document}
\renewcommand{\theequation}{\thesection.\arabic{equation}}
\newcommand{\beq}{\begin{equation}}
\newcommand{\eeq}[1]{\label{#1}\end{equation}}
\newcommand{\ber}{\begin{eqnarray}}
\newcommand{\eer}[1]{\label{#1}\end{eqnarray}}
\begin{titlepage}
\begin{center}

\hfill CERN-TH.7100/93\\
\hfill CPTh-A276.11.93\\
\hfill hep-th/9311185\\

\vskip .5in

{\large \bf  EXACT STRING-THEORY INSTANTONS BY DIMENSIONAL REDUCTION}
\vskip .5in

{\bf C. Bachas} \footnotemark \\

\footnotetext{e-mail address: BACHAS@ORPHEE.POLYTECHNIQUE.FR}

\vskip .1in

{\em Centre de Physique Th\'eorique\\
Ecole Polytechnique \\
 91128 Palaiseau, FRANCE}

\vskip .15in

       and

\vskip .15in

{\bf Elias Kiritsis} \footnote{e-mail address:
KIRITSIS@NXTH08.CERN.CH}\\
\vskip
 .1in

{\em Theory Division, CERN, CH-1211\\
Geneva 23, SWITZERLAND}\\

\vskip .1in

\end{center}

\vskip .4in

\begin{center} {\bf ABSTRACT }
\end{center}
\begin{quotation}\noindent
We identify exact gauge-instanton-like solutions to (super)-string
theory
using the method of dimensional reduction.
We find in particular the
  Polyakov instanton of 3d QED, and a class of
generalized Yang-Mills merons. We
 discuss   their marginal deformations, and show that for the $3d$
instanton they correspond to a dissociation of vector- and axial-
magnetic charges.

\end{quotation}
\vskip 3.0cm
CERN-TH.7100/93 \\
CPTh-A276.11.93\\
November 1993\\
\end{titlepage}
\vfill
\eject
\def\baselinestretch{1.2}
\baselineskip 16 pt
\noindent
\section{Introduction}
\setcounter{equation}{0}

  \ \ \ \ \ A key tool for understanding
 non-perturbative effects in
field theory has been the study of solitons and instantons.
For a given Lagrangian, it is usually straightforward to
 find the classical solutions, calculate their mass
 or action, zero modes e.t.c.. It is of course a harder
  task to derive physical
  consequences, such as the formation
 of fermionic condensates.
In string theory, on the other hand, just finding
the relevant classical solutions has proved to be a non-trivial
exercise \cite{callan}.
 Indeed, most of the solutions of the effective
low-energy Lagrangian are modified by
higher-order in $\alpha'$
corrections, while the known exact conformal models have often
an obscure space-time interpretation.
This letter is meant as an addition to the
{\it world-sheet} versus {\it space-time} dictionnary.
 Our main observation is that by means
of a   dimensional reduction
one can identify certain instantons, merons, monopoles and other
gauge-
(pseudo)particles, with
combinations of WZW and Feigin-Fuchs models,
 and hence
with exact solutions to the $\beta$-function equations, at least to all
orders in the $\alpha '$ expansion.
We also study the exact marginal deformations of these solutions.
These   solutions could   be important for
studying  gaugino condensation and the ensuing
possible breaking of space-time supersymmetry, but
we will not address this issue here.

\def\baselinestretch{1.2}
\baselineskip 16 pt
\noindent
\section{The Instanton of $3d$ QED}

  \ \ \ \ \  Dimensional reduction was used
in the past to reinterpret
gravitational instantons as Kaluza-Klein monopoles \cite{gross}.
In  string-theory it has been used to identify the $SU(1,1)$ $WZW$
model
 with electrovac solutions of
gauged supergravity \cite{abs}, and more recently the $SU(2)$
$WZW$ plus either the $SU(1,1)/U(1)$ GKO coset or Feigin-Fuchs
models, with axionic instantons and their related monopoles \cite{Kh}
as well as with magnetically charged $4d$ black holes \cite{Pol}.
 Let us illustrate the argument
with the closely-related example of the instanton in $3d$ compact
electrodynamics,
 $$ F_{ij} =
 q \epsilon_{ijk}
{ x^{k}\over\vert x \vert^3} \ .\eqno(1)$$
This solves the flat-space (Euclidean) Maxwell equations, and
satisfies
the Bianchi identity everywhere except at the origin of coordinates.
Appended with an appropriate dilaton background,
$$ \Phi = -{1\over 2} log { \vert x\vert^2 } + log({q\over\sqrt{2}})
\
,\eqno(2)$$
 it also solves the
 leading-order $\beta$-function equations,
derived from the effective $3d$ string  Lagrangian
\footnote{A factor-of-two mistake in the $\delta c$ term has
propagated in much of the literature. }
$$ {\cal L}_{eff}^{(3)}  \propto  \int  d^3 x \
 \sqrt{g} \ \Bigl[-R + (\partial \Phi)^2 +
{1\over 4} e^{-2\Phi} F_{ij} F^{ij} - {2 \over 3 \alpha'} \delta c\
e^{2\Phi}\Bigr] \ \ , \eqno(3)$$
with $\delta c = 0$.
 Note that the ``central-charge deficit" $\delta c$,
  depends in general on the details
of the compactification from the critical down
to three dimensions.
 It is straightforward to check that the total
energy-momentum tensor of the above
  monopole- and dilaton-backgrounds
vanishes, so that the $3d$   metric stays flat.

We will now show that the above backgrounds correspond precisely
to an $SU(2)$  WZW plus a Feigin-Fuchs model on the world-sheet.
This $3d$ instanton is therefore an exact solution of bosonic-string
theory, as well as  a reinterpretation of the  much-discussed
(singular) semi-wormhole  \cite{rey}\cite{aben}\cite{callan} .
 Indeed, the corresponding
$\sigma$-model Lagrangian  in conformally flat  coordinates
reads
$$
{\cal L}_{\sigma} = {k\over 4 } \int {d^2\xi \over 4\pi}
 \Bigl[  \d \alpha \db \alpha +\d \beta \db \beta +
\d \gamma \db \gamma  +
   2 cos\beta\    \d \alpha\db\gamma\Bigr]  +
   \int {d^2\xi \over 4\pi}
  \Bigl[  \d {\tilde r} \db {\tilde r} +
 \sqrt{g} R^{(2)}  Q{\tilde r}\Bigr] \ .
\eqno(4)$$
Here ${\tilde r}$ is the Feigin-Fuchs field,
$\alpha\in [0,4\pi]$, $\beta\in[0,\pi] $ and
$\gamma \in
[0,2\pi]$   are Euler angles parametrizing the $SU(2)$ group manifold
 with a non-standard choice of  ranges,
 and unitarity forces $k$ to be a positive integer.
Our notation is as follows:
 $\xi^a$ are the conformally-flat coordinates,
   $z= (\xi^1+i\xi^2)/2$
and we have set the Regge slope $\alpha'=1$.
    The central charge of the above conformal model is
$c = 1 + {3\over 2}Q^2 + 3k/(k+2)$ .
We can read off the $\sigma$-model backgrounds by comparing eq. (4)
to the generic form
$$ {\cal L}_{\sigma} = \int {d^2\xi \over 4\pi}
 \Bigl[  ( G_{IJ} + B_{IJ}) \d X^I \db X^J  +
  \sqrt{g} R^{(2)} \Phi(X) \Bigr] \ , \eqno(5)$$
where $X^I$\  $(I=1,..,4)$ denote collectively the coordinate
fields. To make contact with the $3d$ instanton
 we will now view the angle $\alpha$
as a compact internal coordinate, and
make a Kaluza-Klein decomposition of the
metric and antisymmetric tensors
\footnote{We use lower-case indices to label the space
after dimensional reduction.}:
 $$ \eqalign{ G_{IJ} &= \left( \matrix
{ {\tilde g}_{ij} + v a_i a_j    &v a_i \cr
v a_j &  v \cr}\right) \ \  \cr
B_{IJ} &= \left( \matrix
{ b_{ij} + b_i a_j- a_i b_j & b_i \cr
 -b_j & 0 \cr}\right) \cr}
.\eqno(6)$$
Under this reduction, the reparametrization- and antisymmetric-tensor
invariances in the compact dimension, descend to a vector-
and an axial-$U(1)$ gauge symmetry. The merit of the decomposition
above
is to simplify the corresponding transformation laws.
It is indeed straightforward to check that the $\sigma$-model
action is invariant (up to  boundary terms) under the
following
transformations:
($ \delta a_i = \d_i \Lambda_{vec} $)  and
($ \delta b_i = \d_i \Lambda_{ax}\ ; \
\delta b_{ij} = \Lambda_{ax} F_{ij}(a)$).
Note in passing that the gauge-invariant antisymmetric-tensor
field strength in the reduced dimensions reads
 $H_{ijk} = \d_i b_{jk} - b_i F_{jk}(a) +
{\rm cyclic\  perms}$.

In the case that interests us we can set $v=1$ by rescaling
the internal coordinate $\alpha\to  2\alpha/\sqrt{k}$,
and work with the chiral gauge fields
$A_i = (a_i-b_i)/\sqrt{2}$ and \hfil\break
 ${\bar A}_i = (a_i+b_i)/\sqrt{2}$ .
 We will furthermore go to the
Einstein-frame metric  $ g_{ij} = e^{-2\Phi/(d-2)} {\tilde
g}_{ij}$, in terms of which the effective low-energy
Lagrangian is given by eq. (3).
Defining flat polar coordinates:
$ r \equiv e^{-\Phi({\tilde r})}/Q$, $\theta\equiv\beta, \phi\equiv
\gamma$ we can finally identify the backgrounds of the
$\sigma$-model (4) with the low-energy solution, eqs. (1,2).
Indeed, the background of the
$A_i$ gauge field is precisely the instanton with charge
$$q = \sqrt{k/2}\ , \eqno(7)$$
 while ${\bar A}_i = 0$ and the Einstein-frame
metric is flat provided we choose
$Q^2  = {4\over k} + {\cal O}({1\over k^2}) $.
This choice implies that $\delta c \equiv c-4 $ vanishes
to one-loop order,
in accordance with the effective field-theory
argument. Of course,
the $\sigma$ model, eq. (4),  stays   conformally invariant for
arbitrary values of $Q$. The corresponding
solution is an instanton
in  curved $3d$ space:
$$ ds^2 = dr^2 + {Q^2k\over 4} r^2 d^2\Omega \ ,\eqno(8)$$
where $d^2\Omega$ is here the distance on $S^2$.

Note that the radius of the compact internal dimension
is $R= \sqrt{k}$ \footnote{Had we not identified the corresponding
CFT, we could thus not ascertain its existence from the
effective $3d$ action since a fourth dimension decompactifies
in the weak-field limit.}, so that
 the spectrum of electric charges
is $e_{nm} = (n/(\sqrt{2k})+ m\sqrt{k\over 2}) $ with $n,m$
integers. The ''magnetic" charge, eq. (7), is thus the minimum one
allowed
by the
  Dirac quantization condition:
$2q e \in Z$ for all electric charges $e$ of the theory.
The Dirac quantization condition is furthermore equivalent to
  the unitarity constraint, $k=R^2 \in Z$, of the $WZW$ model
\footnote{This is another facet of an argument originally due to
Rohm and Witten \cite{RW}}.
We could interpret this constraint as saying that
 instantons of a $U(1)$, coming from a purely
holomorphic sector of the string, do not exist
in general due to the presence
of irrationnally-related charges.

The analysis above is a small variation of the one used in refs.
\cite{abs} \cite{Pol} to identify two
 other classes of string-theory solutions.
The first are the electrovac solutions to gauged supergravities,
 discovered by Freedman and Gibbons \cite{FG},
and characterized a priori by both   magnetic and electric
 graviphoton
backgrounds. The solution with vanishing
magnetic field, which  is distinguished
by the existence of
$N=2$ unbroken space-time supersymmetries,
 corresponds precisely to a
  $SU(1,1)$ $WZW$ model plus extra free coordinate fields \cite{abs}.
Replacing the latter by a
$SU(2)$ $WZW$ model, one can in fact reproduce the entire set of the
Freedman-Gibbons solutions, including those that break space-time
supersymmetry with a non-vanishing magnetic field.
The second class of interesting solutions are the
magnetically-charged $4d$ dilatonic black holes \cite{black}.
Limiting cases of these solutions are obtained \cite{Pol} if one adds
to
the $\sigma$ model, eq. (4), a free time-like coordinate $t$, or
if one  replaces the $({\tilde r}, t)$ system by
the exact $2d$ black hole \cite{Witten}.
In \cite{Pol} a left orbifold of the $SU(2)$ WZW model was also
considered.
Its effect is to make the compactification radius of the angle
$\alpha$
in (4) equal to $4\pi/N$, where $N$ is a divisor of $k$.

\newpage

\def\baselinestretch{1.2}
\baselineskip 16 pt
\noindent
\section{ Reduction of general WZW models}

  \ \ \ \ \  Generalizing the above procedure,
one may decompose a $WZW$ model
on an arbitrary group manifold $G$,  into a $H_L\times H_R$
current algebra, a $G/H$ coset manifold treated as
part of non-compact space, and background gauge and
antisymmetric tensor fields \cite{abs}.
We will write the group elements as ${\hat g} =  h(y) g(x)$
where the $y$ coordinates parametrize the $H$-subgroup manifold,
while the
$x$ coordinates parametrize the right coset $G/H$.
 An essential ingredient of the reduction is the
Polyakov-Wiegmann formula
$$ I( h g ) = I(h) + I(g) +   \int {d^2\xi \over
2\pi} tr [
h^{-1}\d h \db g g^{-1}] \ ,\eqno(9)$$
 where $I(g)$ is proportional to the
$WZW$ action of a simple group $G$,
$$ I(g) =   \int {d^2\xi \over 4\pi} tr [ g^{-1}\d g g^{-1}\db g]
-i  \int_{\cal B} {d^3\xi \over 6\pi} \epsilon^{abc} tr [
g^{-1}\d_a g g^{-1}\d_b g g^{-1}\d_c g] \ ,\eqno(10)$$
with
  ${\cal B}$  being as usual a solid ball
 whose  boundary is the Euclidean $2d$  world
sheet.
If the traces are taken in some $R$ representation of the group,
 the correctly normalized $WZW$ action is
$$ S^{WZW}_k = -\kappa I(g) \ \ , \ \
\kappa \equiv {k\over 4} {c_G d_G \over
 {\tilde h}_G c_R d_R} \eqno(11)$$
where
$k\in Z$ is an integer, $d_R$ ($d_G$) and $c_R$ ($c_G$) are
the dimension and quadratic Casimir of the $R$ (adjoint)
 representations
respectively,
 and ${\tilde h}_G$ is the dual Coxeter number:
 ${\tilde h}_G = n$ for $SU(n)$ and ${\tilde h}_G =n-2$ for $SO(n)$ .

 The three terms in eq. (9) above
correspond to a metric and antisymmetric
tensor background on the coset space, a $H_L\times H_R$ current
algebra,
  and a background $H_L$ gauge field .
In order to read off these backgrounds we will use again
 gauge invariance as
a guide.
Our starting point is the vacuum consisting of flat space-time
($x^\mu$)
plus the $H_L\times H_R$ current algebra.
Under a gauge transformation
$\Bigl(\upsilon(x), {\bar\upsilon}(x)\Bigr)$
, the   corresponding
left- and right-gauge fields  transform as follows
$$ \eqalign{
  A_{\mu}\to &\upsilon  ( A_{\mu} -
\upsilon^{-1}\d_{\mu} \upsilon) \upsilon^{-1}   \cr
  {\bar A}_{\mu}\to &{\bar \upsilon}
 ( {\bar A}_{\mu} - {\bar \upsilon}^{-1}\d_{\mu} {\bar \upsilon})
 {\bar \upsilon}^{-1} \ \ , \cr}
\eqno(12)
$$
where with our conventions   the gauge-field
strengths
read: $ F_{\mu\nu} = \d_\mu A_\nu - \d_\nu A_\mu + [A_\mu,A_\nu]$,
with a similar expression for ${\bar F}_{\mu\nu}$.
Next let us define the Lie-algebra valued vector field
on the world sheet \footnote{To avoid confusion we stress that
$h(\xi)$ and $x^\mu(\xi)$ are string coordinates, while
$\upsilon(x)$ and ${\bar \upsilon}(x)$ stand for target-space gauge
transformations.},
$$
J_a \equiv h^{-1}\d_a h +( h^{-1} {\bar A}_\mu h - A_{\mu} )
\d_a x^\mu  \ \ , \eqno(13a)$$
and its conjugate
$$ {\bar J}_a \equiv h J_a h^{-1}\ \ . \eqno(13b)$$
Under a target-space gauge transformation  and
a simultaneous change of string coordinates,
$$ h \to {\bar \upsilon}(x)\  h\ \upsilon^{-1}(x) \ \ ,\eqno(14)$$
these vector fields transform homogeneously:
$$ J_a \to \upsilon  J_a \upsilon^{-1} \ \ ; \ \
{\bar J}_a \to {\bar \upsilon}
 {\bar J}_a {\bar \upsilon}^{-1} \ \ .\eqno(15)$$
We may then write the following $\sigma$-model action to describe
the interaction of a string with arbitrary
metric , antisymmetric-tensor and gauge-field
backgrounds:
$$\eqalign{ S = \int {d^2\xi\over 4\pi}{\tilde g}_{\mu\nu}&
\d x^\mu \db x^\nu
 + i  \int_{\cal B} {d^3\xi \over 12\pi} \epsilon^{abc}
{\hat H}_{\mu\nu\rho}\d_a x^\mu \d_b x^\nu \d_c x^\rho
 - \kappa  \int {d^2\xi\over 4\pi}
 tr (J_z  J_{\bar z})   \cr
+ &i\kappa \int_{\cal B} {d^3\xi \over 6\pi} \epsilon^{abc} \Bigl[
tr(
J_a J_b J_c) -
{3\over 2} tr\Bigl( F_{\mu\nu}  J_c + {\bar F}_{\mu\nu}  {\bar J}_c
\Bigr) \d_a x^\mu \d_b x^\nu
 \Bigr] \ . \cr}
   \eqno(16)$$
Assuming ${\tilde g}_{\mu\nu}$ and ${\hat H}_{\mu\nu\rho}$ do
not transform,
this action is manifestly gauge-invariant. Furthermore, it   reduces
to  $-\kappa I(h) + \int (\d x)^2$ for vanishing background
deformations.
To prove however that it is well-defined, we must still
show that the $3d$ integrand is a total divergence.
This requirement is what fixed in particular the relative
coefficient of the last   term of the action.
After a tedious but
straightforward calculation, one may   put eq. (16)
in the form:
$$\eqalign{ S = -\kappa I(h) + \int {d^2\xi\over 4\pi}
\Bigl[ {\tilde g}_{\mu\nu}&
 - \kappa \ tr ( A_\mu A_\nu
+ {\bar A}_\mu {\bar A}_\nu - 2 {\bar A}_\mu h A_\nu h^{-1} ) \Bigr]
\d x^\mu \db x^\nu
\cr
 + \kappa & \int {d^2\xi\over 2\pi}\Bigl[
 tr(A_\mu h^{-1}\d h ) \db x^\mu -
tr({\bar A}_\mu \db h
h^{-1})
\d x^\mu \Bigr]
\cr
 + i  \int_{\cal B} &{d^3\xi \over 12\pi} \epsilon^{abc}
\Bigl[ {\hat H}_{\mu\nu\rho} + \kappa CS_{\mu\nu\rho}(A)
- \kappa CS_{\mu\nu\rho}({\bar A}) \Bigr]
\d_a x^\mu \d_b x^\nu \d_c x^\rho
 \ \ ,  \cr}
  \eqno(17)$$
where the Chern-Simmons three-form is
$$ CS_{\mu\nu\rho}(A) \equiv tr( A_\mu F_{\nu\rho} - {1\over 3} A_\mu
[A_\nu, A_\rho] + cyclic \  perms) \ . \eqno(18)$$
It is now clear that for the action to be well-defined, we
must demand that the three-form in square brackets be exact:
$$ {\hat H}_{\mu\nu\rho} + \kappa CS_{\mu\nu\rho}(A)
- \kappa CS_{\mu\nu\rho}({\bar A}) \equiv H_{\mu\nu\rho} \equiv
\d_\mu b_{\nu\rho} + cyclic\  perms \ \ . \eqno(19)$$
We have thus rederived from the
$\sigma$-model  the Green-Schwarz mechanism of anomaly
cancellation: the antisymmetric-tensor field $b_{\mu\nu}$
must transform so as to make the generalized field strength
${\hat H}_{\mu\nu\rho}$, through which it enters
in the effective low-energy Lagrangian,  gauge-invariant.
A similar derivation was given before for the heterotic string
by Hull and Witten
\cite{Hull}. In their case the Chern-Simmons
transformation law came from the gauge anomaly of the fermions, while
in our case it comes from the WZW term of the action.

Eq. (17) is our basic formula, which allows the identification of
backgrounds for any group-manifold
 compactification\footnote{Explicit formulae for
generic dimensional reductions can be found in refs. \cite{Scherk}.}.
There is
  in fact an extra gauge-invariant term that we could have added
  to the $\sigma$-model action. It is
$$ S' = \int {d^2\xi\over 4\pi} \Phi_{\alpha {\bar \beta}}(x)
tr(J_z T^{\alpha}) tr({\bar J}_{\bar z} T^{\bar \beta}) \,  \eqno(20)
$$
where $T^\alpha$ are the group generators, and the scalar
$\Phi_{\alpha {\bar\beta}}$ transforms in the $(adj,adj)$
representation of the gauge group. This term
  corresponds to space-dependent deformations
  of the metric and antisymmetric
tensor in the compact directions.
Comparing the Polyakov-Wiegmann formula
 with the generic form (17) and (20), one sees
immediately
that for the right-coset
reductions of $WZW$ models $\Phi_{\alpha{\bar \beta}} =
{\bar A}_\mu
=0$. The only non-trivial backgrounds are therefore
$  A_\mu$, $b_{\mu\nu}$ and the space-time metric which,
as can be verified easily,   is always the metric
of the symmetric space $G/H$. Let us however point out that other
coset reductions are sometimes possible. Thus,
 if $H\simeq H_1\times H_2$
is not semisimple, the decomposition ${\hat g} = h_1 g h_2$ will
lead to both $\Phi_{\alpha{\bar\beta}}$ and ${\bar A}_\mu$
backgrounds.

\def\baselinestretch{1.2}
\baselineskip 16 pt
\noindent
\section{ Yang-Mills merons}

\ \ \ \ \ We will now apply this procedure
 to the simplest example with non-abelian
gauge fields, namely the $SU(2)_{k_1}
\times SU(2)_{k_2}$ $WZW$ model reduced so that a diagonal subgroup
  $H$ is internal space \footnote{
Had we chosen H as one
of the two SU(2) groups, the resulting gauge field background
would be a pure gauge.}.
Let $(g_1, g_2) \equiv (h, gh)$ be the corresponding decomposition
of the
$SU(2)\times SU(2)$ group manifold, where $h$ and $g$ are independent
$2\times2$ unitary matrices. We will parametrize these latter
as follows:
  $h= y^0 {\bf 1} + i {\vec \s}\cdot
{\vec
y}$
and $g = w^0 {\bf 1} + i {\vec \s}\cdot{\vec w}$,
where $
   (y^0)^2 + {\vec y}\cdot{\vec y}= (w^0)^2+{\vec w}\cdot {\vec w}=
1$,
and ${\s^i}$ are the   Pauli matrices normalized
 so that $\s^i\s^j=\delta_{ij}+
i\epsilon_{ijl}\s^l$.
 Using the Polyakov-Wiegmann formula, we can write the
WZW action of the model:
$$- S_{WZW} = {k_1+k_2\over 2} I(h) + {k_2\over 2}
 I(g) + k_2 \int {d^2\xi\over 4\pi}
tr[h^{-1}\d h
\db g g^{-1}]\ . \eqno(21)$$
Comparing with eq. (17), we can   read off the following
non-vanishing
 backgrounds:
$$ {\tilde g}_{ij} = {k_1 k_2\over k_1+k_2} \left( \delta_{ij} + {w^i
w^j\over
(w^0)^2}\right)\ ,
 $$
$$ H_{ijl} = 2 k_2\  \epsilon_{ijl}/ w^0  \ ,
\eqno(22)$$
$$ A_i = - {k_2\over k_1 + k_2}
 \partial_ig(w)\ g(w)^{-1}\
  . $$
Note that
the metric and antisymmetric-tensor field strength   are
proportional,
respectively, to
the metric and volume form  on $S^3$, as for a simple $SU(2)$ $WZW$
model.
The radius square of the sphere need not, however, here be an
integer.

As in the case of the $3d$ instanton, we can again add
  an extra
Feigin-Fuchs coordinate $r$ ,
with background charge $$ Q^2 =
{4(k_1 + k_2)\over k_1 k_2} \ , \eqno(23)$$
so as to render the $4d$ Einstein-frame metric flat.
Going to the
  flat Cartesian coordinates
$$ x^\mu = {2\over Q} e^{-\Phi/2}\ w^\mu  \ ,
\eqno(24)$$
and trading the antisymmetric tensor for a pseudo-scalar axion field
through the duality transformation
$$ {\hat H}_{\mu\nu\rho} \equiv  H_{\mu\nu\rho} + {k_1+k_2\over 2}
CS_{\mu\nu\rho}(A) =
\epsilon_{\mu\nu\rho\sigma}e^{2\Phi}\d^{\sigma}b\ ,\eqno(25)$$
we obtain:
 $$\eqalign{ &g_{\mu\nu}=\delta_{\mu\nu}\cr
\Phi &= -log\vert x\vert ^2 - log{Q^2\over 4} \cr
 b&= {k_1-k_2\over k_1 k_2} \vert x\vert^2\  , \cr} \eqno(26)$$
while the gauge-field in these coordinates becomes
 $$ A_\mu = -{i\over 2} \s^a  A_{\mu}^a \ \ , \ \
A_{\mu}^a = {2k_2\over k_1+k_2} {\eta^a _{\mu\nu} x^\nu \over \vert
x\vert^2} \ \ .\eqno(27)$$
where
  $\eta^a_{\mu\nu}$ are the well-known $\eta$-symbols
describing the projection of
SO(4) into a left SU(2) subgroup
\cite{tHooft}.

It   is straightforward to check that the above backgrounds solve
the field equations derived from
   the effective $4d$ Euclidean Lagrangian
$$\eqalign{
 {\cal L} =  \int  d^4x \sqrt{g} \Bigl[ -R +{1\over 2} (\partial
\Phi)^2
- & {1\over 2} e^{2\Phi} (\partial b)^2
+ {1\over 4g^2} e^{- \Phi} F_{\mu\nu}^a  F^{a\mu\nu} \cr
 + &{1\over 4g^2} b  \ F^{a\mu\nu} \ ^*F_{\mu\nu}^a
 - {2\over 3 }\delta c\  e^{\Phi} \Bigr] \  . \cr} \eqno(28)$$
Here
$^{*}F^{a}_{\mu\nu}={1\over
2}{\epsilon_{\mu\nu}}^{\rho\sigma}F^{a}_{\rho\sigma} $, \
$ g^2 = 2/(k_1+k_2)$ is the $SU(2)$ coupling constant
for zero value of the dilaton, and the central-charge deficit
is fixed from the dilaton
equation to be :
$$  \delta c = 3g^2 =
6/(k_1+k_2)\ .
\eqno(29)$$
This is of course consistent
with  the conformal-field-theory result in the
 $k_1,k_2\rightarrow\infty$ limit.
Note that a rotation of the axion, $b\to ib$, makes
 its kinetic term positive definite
and the $F\ ^* F$ term purely imaginary. One should however keep in
mind
  that the Euclidean
functional integral must be defined in terms of the fundamental
field $b_{\mu\nu}$, so that a priori the relevant
 saddle points of (28) are real.
Note also that
the constant part of the dilaton,
 which  was a free parameter of the conformal
model, has been here fixed by requiring that the
space metric be $\delta_{\mu\nu}$.

We may choose the remaining two free parameters to be
 the gauge coupling
  $g$,  and $\lambda= k_2/(k_1+k_2)$, so that the
 gauge-field background reads
$$A_{\mu}^a = 2\lambda {\eta^a _{\mu\nu} x^\nu \over \vert
x\vert^2}\ . \eqno(30)$$
This is a generalization of a well-known solution of
pure
Yang Mills theory, known as the {\it meron} \cite{Fubini}.
The pure Yang-Mills meron has
$\lambda = {1\over 2}$,
 and is characterized by a topological charge $q={1\over 2}$
residing at the origin of coordinate space. We can obtain the
  charge of our generalized merons by integrating
the Chern-Simmons form on a three-sphere around the
origin, with the result
$$ q(\lambda) \equiv  -{1\over 48\pi^2}
\int_{S^3}\ \epsilon_{\mu\nu\rho\sigma}
CS^{\mu\nu\rho} {\hat x}^\sigma \ d^3x \ = \  \lambda^2
(3- 2\lambda)\ . \eqno(31)$$
This calculation   appears to be incompatible with
the axion equation $\d_\mu (e^{2\Phi} \d^\mu b)+{1\over 4g^2} F \
^*F
= 0$, at the origin. Both terms are proportional to a
$\delta$-function, but the equation yields
$\lambda(\lambda-1)(2\lambda-1) +
  \lambda^2 (3-2\lambda) \equiv \lambda = 0$, which is not satisfied

in general. This is however, not surprising: although our
solution is exact, the validity of the low-energy equations
of motion clearly breaks down at $x=0$.

\setcounter{footnote}{0}

Demanding unitarity of the $WZW$ model, imposes a quantization
condition
on $\lambda$ and on the corresponding topological charge. If we
let $k\equiv k_1+k_2$ be the level of the (internal) current algebra,
 and $n\equiv
k_1-k_2$, then the spectrum of allowed charges is
$$ q_n = {1\over 4} (1-{n\over k})^2 (2+{n\over k}) \ \ \ \ \
(n = -k, ... , k) . \eqno(32)$$
Note that the higher the level $k$ of the current algebra, the richer
the spectrum of merons. Note also that the conjugation
 $g\rightarrow g^{\dag}$
transforms the meron to an anti-meron, and that a meron with
parameter
$\lambda$ transforms under the (singular)
 gauge transformation $h\rightarrow h g^{\dag}$ to an
antimeron with parameter
$1-\lambda$. A simple consequence of this fact is that
$q(\lambda) + q(1-\lambda) =1$. Note finally that for $\lambda=1$,
$A_\mu$
is the field of an instanton \cite{Poly}, but in the zero-size limit
in which it is a pure
  (singular) gauge. The string solution has, however,
  non-trivial
dilaton and axion backgrounds, and is in fact the well-known
semi-wormhole \cite{rey} \cite{aben} \cite{callan}.

Despite much discussion in the literature, the physical
interpretation
of merons remains still unclear. They suffer  from a singularity
at $x=0$, and   an infrared-divergent action, but are in this respet
similar to  vortices in the $2d$
XY model.
This analogy suggests that they could be instrumental for
confinement
in four dimensions \cite{CDG}. They have been also interpreted as
half instantons and as short-lived monopoles. In pure Yang-Mills
theory
they survive unchanged in any conformally flat isotropic space
 \footnote{ Callan
and Wilczek \cite{CW} have made the interesting suggestion of using
such
a negative-curvature space
  as a gauge-invariant infrared regulator. Let us note that this role
could be also played by a   dilaton background, that
 could suppress strong
fluctuations outside some finite region.}.
By relaxing the condition (23) on the background
 charge of the dilaton, we
can also extend the string-theory merons to any conformally-flat
space
of the form $g_{\mu\nu} \propto
\vert x \vert ^{2\alpha -2} \delta_{\mu\nu}$.
The axion and dilaton backgrounds in this case
read $ b \propto e^{-\phi} \propto
\vert x \vert ^{2\alpha}$, where at the level of
 the $\beta$-function
equations, $\alpha$  is an arbitrary continuous parameter of the
solution.

Before closing this section let us point out  that higher-dimensional
generalizations of the $3d$ QED instanton, and
the $4d$ Yang-Mills meron can be obtained easily through the
dimensional
reduction $SO(N+1)\to SO(N)$. And that
  another class of CFTs, which contains current algebra
and can be thus used for non-abelian compactifications, are
  gauged WZW models $G/H$ where $H$ is not maximal.
 It can be shown
\cite{k4} that these models have chiral $H'$ currents,
for any subgroup $H^\prime$
commuting with $H$.
Dimensional reduction can be thus performed also here but we will not
elaborate this case further.

\def\baselinestretch{1.2}
\baselineskip 16 pt
\noindent
\section{ Deformations}

\ \ \ \ \ The conformal models
  of the previous sections  have continuous
deformations or moduli. From the lower-dimensional perspective, these
  deformations   will, in general, involve massive Kaluza-Klein
excitations. One way to ensure that
 only massless lower-dimensional
backgrounds are present, is to consider perturbations that respect
the $H_R$ isometry of the $WZW$ model.
 Changing the background charge
of the Feigin-Fuchs field is one trivial example of this type. Here
we want to discuss deformations of the $WZW$ model itself.
Deformations which mix the Feigin-Fuchs
and $WZW$ coordinates would be very interesting,
since they could modify non-trivially
the radial dependence of
the backgrounds, but we do not know how to handle them analytically
at present.

The generic \footnote{For special values of the level
there may
be more marginal directions.}  exactly marginal perturbation of the
WZW model is generated by the Cartan currents, $S_{pert}\sim
C_{\alpha\beta}\int J^{\alpha}\bar J^{\beta}$.
Such deformations a priori break the local $G_{L}\times G_{R}$
symmetry down to $U(1)^{\rm rank_G}_{L}\times U(1)^{\rm rank_G}_{R}$,
but we can
chose to leave a larger subgroup of
the
current algebra unbroken. The corresponding deformed $\sigma$-model
action is known explicitly \cite{hs,gk} only for
    perturbations  $ S_{pert} \sim \int J\bar J$,
where $J$ corresponds to a single Cartan generator $T$ of the group.
Let us
 normalize this (antihermitean) generator so
that $tr(TT)= -{1\over 2}$, and
parametrize the group manifold as follows:
   $g=e^{T \tbb}\gt(w) e^{T \ta}$, with $w^I$  the
$({\sl dim}G -2)$ remaining coordinates.
 Using the Polyakov-Wiegmann formula
we can write the $WZW$ action in the
 form:
 $$-\kappa I(g) = -\kappa I({\tilde g})
 +
\frac{\kappa}{2}\int {d^2\xi\over 4\pi}  \Bigl(\d\ta \db\ta + \d\tbb
\db\tbb \Bigr) + \ \ \ \ \ \ \ \ \ \ \ $$
$$ + {\kappa\over 2} \int {d^2\xi\over 2\pi}
 \Bigl(  \S (w) \d\tbb \db\ta
 + \G_I^1(w) \d w^I\db\ta +  \G_I^2(w) \d\tbb \db
w^I\Bigr)\eqno(33)$$
where  here
$$\G^{1}_{I}(w)\d w^{I}= -2 tr(T \db\gt
\gt^{-1})  \ ,$$
$$ \G^{2}_{I}(w)\db w^{I}= -2 tr(T \gt^{-1}\d
\gt) \
 ,\eqno(34)$$
 $$\S(w) =  -2 tr({\tilde g}^{-1} T {\tilde g} T) \ . $$
This action is manifestly invariant under
$\ta\to\ta+\epsilon (z)$ and $\tbb\to\tbb+\bar \epsilon(\bar
z)$. The corresponding
 left
and right chiral currents are
$$J={\kappa\over 2}\left(\d\ta+\S(w)\d\tbb+\G^{1}_{I}\d
w^{I}\right)\eqno(35a)$$
and
$$\bar J={\kappa\over 2}\left(\db\tbb+\S(w)\db\ta+\G^{2}_{I}\db
w^{I}\right) \ .\eqno(35b)$$
The marginal perturbation  $S_{pert}\sim \int
J\bar J$ then leads to the following continuous family
of conformally-invariant
$\sigma$-models
  \cite{gk}:

$$S(\kappa, \zeta) =
 -\kappa I({\tilde g}) + {\kappa\over
2}\int {d^2\xi\over  4\pi} \left[\d\phi_{1}\db\phi_{1}+
\d\phi_{2}\db\phi_{2}+
2{1+\S-\zeta^2(1-\S)\over 1+
\S+\zeta^2(1-\S)}\d\phi_{2}\db\phi_{1}+\right.$$
$$\left.+{2\zeta\over 1+\S+\zeta^2(1-\S)}\left(\G^{1}_{I}\d
w^{I}\db\phi_{1}+\G^{2}_{I}\d\phi_{2}\db w^{I}\right)+ {{1\over 2}
 (\zeta^2-1)\G^{1}_{I}\G^{2}_{J}\over 1+\S+\zeta^2(1-\S)}\d
w^{I}\db w^{J} \right]\eqno(36)$$
where
$$\phi_{1}={1\over 2}\left(\zeta+{1\over \zeta}\right)\ta+{1\over
2}\left(\zeta-{1\over \zeta}\right)\tbb\;\;,\;\;\phi_{2}={1\over
2}\left(\zeta-{1\over \zeta}\right)\ta+{1\over 2}\left(\zeta+{1\over
\zeta}\right)\tbb\ \ , \eqno(37)$$
and there is also a non-trivial dilaton background
$$\Phi(w)=-\log\left[\zeta(1-\S)+(1+\S)/\zeta\right] \ .\eqno(38)$$
Clearly, the
  original $U(1)_{L}\times U(1)_{R}$ symmetry of the unperturbed
model
survives  for arbitrary $\zeta$,
 even though the currents themselves get modified:
$$J(\zeta)={\kappa\over 2}\left(\d\phi_{1}+{1+\S-\zeta^2(1-\S)\over
1+\S+\zeta^2(1-\S)}\d\phi_{2}+
{\zeta\over 1+\S+\zeta^2(1-\S)}\G^{1}_{I}\d w^{I}\right)\
,\eqno(39a)$$
$$\bar J(\zeta)={\kappa\over 2}\left(\db
\phi_{2}+{1+\S-\zeta^2(1-\S)\over
1+\S+\zeta^2(1-\S)}\db\phi_{1}
+{\zeta\over 1+\S+\zeta^2(1-\S)}\G^{2}_{I}\db w^{I}\right)\
.\eqno(39b)$$
It can
 be verified easily that $S(\kappa,\zeta)$ reduces to the
$WZW$ action at $\zeta=1$, and that furthermore
 $$S(\kappa,\zeta+\delta \zeta)=
S(\kappa,\zeta)+{2\delta \zeta\over \pi \kappa\zeta}\int J(\zeta)\bar
J(\zeta)+{\cal{O}}(\delta \zeta^2)\ ,\eqno(40)$$
 as claimed.

Let us restrict now our attention to the simplest case of $SU(2)$.
Comparing eqs. (4) and (33) we can identify
  $\ta$, $ \tbb$ and $w$ with the Euler angles
  $\alpha$, $\gamma$ and $\beta$, so that
$\G^{1}=\G^{2}=0$, and $\S=\cos\beta$. The action of the deformed
model reads
$$S(k, \zeta) =
 { k\over
4}\int {d^2\xi\over  4\pi} \left[\d\phi_{1}
\db\phi_{1}+\d\phi_{2}\db\phi_{2}+ \d\beta
\db\beta + 2 F(\beta)
\d\phi_{2}\db\phi_{1}\right] \ , \eqno(41)$$
with
$$ F \equiv  {1+\cos\beta-\zeta^2(1-\cos\beta)\over
1+\cos\beta+\zeta^2(1-\cos\beta)}\ \ .\eqno(42)$$
One can think of
the $SU(2)$ $WZW$ model   as constructed from a free
boson and order-$k$ parafermions, with the actions of a discrete
$Z_k$ symmetry identified. The above family of deformed models
corresponds
in this language to the
continuous variation of the radius of the free boson \cite{gk}.

In order to read off from the   action (41) the deformation of the
$3d$ instanton background of section 2, we must still decide what
combinations of $\phi_1$ and $\phi_2$ will stand for the internal
compact dimension, and for the polar angle of ''real space".
Since, up to a gauge transformation,  these must
 be independently periodic, we should a
priori  identify them with the
Euler angles $\alpha$ and $\gamma$, related to $\phi_1$ and $\phi_2$
through eq. (37). We will work for convenience with the rescaled
angle
$\sqrt{k}\alpha/2 \in [0, 2\pi\sqrt{k})$.
We will furthermore drop the radial Feigin-Fuchs coordinate and the
rescaling
to the Einstein-frame metric, since they will play no role in the
discussion that follows.
Comparing (41) with eqs. (5) and (6) we then find the following
backgrounds
for the scalar, the vector and axial gauge fields, and the $2d$
metric:
$$ v =   { \zeta^2(1+cos\beta) + (1-cos\beta) \over
(1+cos\beta) + \zeta^2 (1-cos\beta)} \ \ , \eqno(43)$$
$$ a_\gamma =  \sqrt{k}  \ {1+cos\beta \over (1+cos\beta) +
(1-cos\beta)/\zeta^2} \ \ , \eqno(44)$$
$$ b_\gamma = \sqrt{k}  \ {1+cos\beta \over (1+cos\beta) +
\zeta^2 (1-cos\beta)} \ \ , \eqno(45)$$
and
$$ ds^2 =  {k\over 4} \Bigl[ d\beta^2 + {1-F^2\over v} d\gamma^2
\Bigr] \ .
\eqno(46)$$
We have here used gauge transformations to make $a_\gamma$ and
$b_\gamma$
vanish at the south pole.
Using the expansions $F\simeq 1 - \zeta^2\beta^2/2$ and
$F\simeq -1 + \zeta^2 (\pi-\beta)^2/2$ near the north and south poles
of
the deformed sphere, one can verify easily that the metric (46) has
no
conic singularities at these points. Furthermore $a_\gamma \simeq
b_\gamma
\simeq \sqrt{k}$
near the north pole, so that their Dirac string singularities are
indeed  unobservable \footnote{ Another way of saying this for the
axial
field,
is that the deformed volume form  $F_{\beta\gamma}(b) d\alpha d\beta
d\gamma$
 has a
$\zeta$-independent normalization.}.

It is in fact possible to demonstrate that the  choice of
periodicities for the angles $\phi_1$ and $\phi_2$, can be uniquely
fixed
by requiring the absence of conic and observable Dirac-string
singularities
for the $\sigma$-model backgrounds.
Indeed, suppose
$$\phi_1 = a_1{\tilde\alpha} + c_1{\tilde\gamma} \ \ ; \ \
\phi_2 = a_2{\tilde\alpha} + c_2{\tilde\gamma} \eqno(47)$$
were some arbitrary linear
combinations of the compact coordinate
 ${\tilde\alpha}\equiv {\tilde\alpha}+4\pi$
and the polar angle ${\tilde\gamma}\equiv{\tilde\gamma}+2\pi$.
After some straightforward but lengthy algebra, one finds that the
$2d$
metric is free of conic singularities provided
$$ {c_1+c_2\over (a_1+a_2)^2}(a_1^2c_2+a_2^2c_1) -
 c_1c_2 = {1\over \zeta^2} \ ,
\eqno(48)$$
and
$$ {c_1-c_2\over (a_1-a_2)^2}(a_2^2c_1-a_1^2c_2) + c_1c_2 =  \zeta^2
\ ,
 \eqno(49)$$
 and that, assuming
continuity in $\zeta$,
  the Dirac quantization conditions for the gauge fields  read
$$ {c_1+c_2\over a_1+a_2} - {c_1-c_2\over a_1-a_2} = 2 \ ,\eqno(50)$$
and
$$ a_2c_1-a_1c_2 = -1 \ . \eqno(51)$$
Modulo a gauge transformation ${\tilde\alpha}\rightarrow
{\tilde\alpha}+\epsilon{\tilde\gamma}$, the unique
solution to the above
constraints is given by the linear combinations (37),
 where ${\tilde\alpha}$
and ${\tilde\gamma}$ are identified with $\ta$ and  $\tbb$, i.e.
precisely
with the Euler angles.

 The meaning of this result is the following:  the perturbative
$\beta$-function equations are satisfied for any choice of the
periodicity-
lattice in the $(\phi_1,\phi_2)$ plane. Non-perturbative effects on
the world-sheet will however break this continuous degeneracy, and it
is reasonable to assume that the allowed periodicities should be the
ones corresponding
to smooth backgrounds \footnote{Orbifold singularities teach us
however
that there could be exceptions to this rule.
The reader may also object that the solutions of the previous sections
have
singularities at the origin. These can be attributed to the
Feigin-Fuchs
model whose consistency as a full-fledged CFT remains indeed to
demonstrate.
There could in particular exist quantization conditions for
 the background charge $Q$. Thus the $3d$ instanton and $4d$ merons are
strictly speaking solutions to all orders in the $\alpha^\prime$
expansion.
}.
  From the string-field-theory point of view, the
degeneracy above corresponds to continuous global symmetries of the
effective low-energy action,
which are broken by the massive Kaluza-Klein excitations.
To be more precise, consider the effective action of a string
  compactified on a circle from
three down to two dimensions:
$$ S_{eff} \propto  \int d^2x \ e^{-\Phi}\sqrt{gv}
 \Bigl[ -R - (\partial\Phi)^2  +
{2\over \sqrt{v}} \triangle \sqrt{v}
  +{v\over 4} F_{\mu\nu}(a)^2
+ {1\over 4v} F_{\mu\nu}(b)^2 \Bigr]\ . \eqno(52)$$
The field equations derived from this action have two manifest
scaling symmetries: ({\it i}) \
$v\to \lambda^2 v$,
$a_\mu\to {1\over\lambda}a_\mu$ and  $b_\mu\to\lambda b_\mu$, and
({\it ii})\  $g_{\mu\nu}\to \lambda^2 g_{\mu\nu}$, $a_\mu\to \lambda
a_\mu$
and $b_\mu\to\lambda b_\mu$, which can be used to transform
one solution to another.
 The
first transformation
  changes the
size of the compact dimension, and boosts the vector and axial
magnetic
charges
of the soliton.
If we were dealing with the ''vacuum", this would of course be an
exact marginal deformation.
In our case, on the other hand, only  combinations of   ({\it i})
and ({\it ii}) which amount to
 discrete rescalings of $k$ (so that $k\in Z$ always)
 take us from one acceptable solution to another. We can say
that the presence of ''matter" has removed the continuous degeneracy
of the
vacuum! Strictly speaking we have, however, traded this degeneracy
for
a new continuous parameter, $\zeta$.
The nature of the $\zeta$ deformation is  however different. At
$\zeta=1$
the vector and axial magnetic fluxes are distributed uniformly over
the
sphere. As $\zeta$ is being increased, the total magnetic charges do
not
change,
but their fluxes  start concentrating
 near the south and north poles, respectively. At the same time they
deform in their vicinity the radius field so as to minimize their
energy.
We can thus describe this deformation as the {\it dissociation},
rather
than
boosting of magnetic charges.
These effects are very much reminiscent of stringy bags \cite{BT},
and
it is intriguing to explore if the analogy can be pushed further.

\def\baselinestretch{1.2}
\baselineskip 16 pt
\noindent
\section{Concluding Remarks}

\ \ \ \ One of the motivations for this work is the hope of getting
some
handle
on the most popular scenario for supersymmetry breaking in
superstrings,
which relies on the condensation of gauginos \cite{gaugi}. In field
theory
one can compute condensates in terms of fermionic zero modes in
instanton
backgrounds.
Identifying exact gauge-instanton-like solutions of the
$\beta$-function
equations, is thus
  a natural first step before attacking the problem at the
string-theory level.
The solutions presented in this paper can be extended easily to the
type-II and heterotic-string case. The basic ingredient is the
 N=1 supersymmetric version of the WZW
model
\cite{Divec}, which contains $\sl dimG$ free fermions in addition to
the
bosonic
group coordinates. In the case of the heterotic string, the
left-moving
fermions of the super-$WZW$ model can be interpreted as part of the
32 fermions that generate the gauge group at the critical dimension,
or
more
generally as part of the
 $(c, {\bar c}) = (26-d, 15-{3\over 2}d)$ conformal field theory,
after compactification to $d$ dimensions.
We should also point out that some of these solutions admit extended
world-sheet supersymmetries. This is for instance the case for the
semi-wormhole or $U(1)$ instanton of section 2, if one chooses the
Feigin-Fuchs charge $Q$ so as to make the central charge $c=4$
exactly
\cite{callan}\cite{Kounnas}.

 In conclusion let us, however, note that
  the hard questions are still ahead of us. First, the exact
solutions
found here have no scale, and are neither asymptotically flat nor
smooth
at the origin. It would be particularly intriguing to have exact
deformations that mix the Feigin-Fuchs and $WZW$ models and introduce
a
scale.
Second, we do not know the (analog of) the
action, collective coordinates and
 normalizable fermionic zero modes in these backgrounds.
 And more generally how to use
such solutions in a ''string-field" functional integral.
We hope some progress will be made on these issues in the near
future.

\vskip 1cm
\noindent
{\bf Acknowledgments} \\
We aknowledge a useful conversation with A. Chakrabarti.
  C.B. thanks the theory division at CERN for hospitality, and
aknowledges support from EEC contracts SC1*-CP92-0792
and CHRX-CT93-0340. E.K. aknowledges the hospitality of the Centre de
Physique Th\'eorique of the Ecole Polytechnique.

\end{document}